\documentclass[12pt,preprint]{aastex}

\usepackage{color}
\usepackage{natbib}
\usepackage{amsmath}
\tightenlines
\slugcomment{}

\newcommand \abs        {{\rm abs}}
\newcommand \alphaG     {\alpha_{\rm G}}

\newcommand \bB         {{\bf B}}

\newcommand \bE         {{\bf E}}

\newcommand \beq        {\begin{equation}}
\newcommand \beqa	{\begin{eqnarray}}

\newcommand \cm         {\,{\rm cm}}

\newcommand \eeq	{\end{equation}}
\newcommand \eeqa	{\end{eqnarray}}
\newcommand \erg	{\,{\rm ergs}}

\newcommand \GHz        {\,{\rm GHz}}
\newcommand \gtsim	{\gtrsim}		 

\newcommand \Ha 	{{\rm H}}

\newcommand \HH	        {{\rm H}_2}

\newcommand \Jy	{\,{\rm Jy}}

\newcommand \K  	{\,{\rm K}}
\newcommand \kms	{\,{\rm km~s}^{-1}}
\newcommand \kpc	{\,{\rm kpc}}
\newcommand \Lsol	{L_{\odot}}
\newcommand \ltsim	{\lesssim}		 

\newcommand \Msol	{M_{\odot}}

\newcommand \nH         {n_{\rm H}}

\newcommand \qpah       {q_{\rm PAH}}
\newcommand \s	        {\,{\rm s}}
\newcommand \sr  	{\,{\rm sr}}

\newcommand \Umax       {U_{\rm max}}
\newcommand \Umin       {U_{\rm min}}

\newcommand \mm         {\,{\rm mm}}
\newcommand \nm         {\,{\rm nm}}

\newcommand{\oldtext}[1]{}

\countdef\decade=200
\advance\decade by \year
\countdef\hours=201
\advance\hours by \time
\divide\hours by 60
\countdef\mins=202
\advance\mins by \hours
\multiply\mins by 60
\multiply\hours by 100
\countdef\miltime=203
\advance\miltime by \hours
\advance\miltime by \time
\advance\miltime by -\mins

\begin{document}

\title{%
 	The Submm and mm Excess of the SMC: \\
        Magnetic Dipole Emission 
        from Magnetic Nanoparticles?
	}

\author{B.T. Draine and Brandon Hensley}
\affil{Princeton University Observatory, Peyton Hall, Princeton,
       NJ 08544; {\tt draine@astro.princeton.edu}}

\begin{abstract}
The Small Magellanic Cloud (SMC)
has surprisingly strong submm and mm-wavelength emission
that is inconsistent with standard dust models, including
those with emission from spinning dust.
Here we show that the emission from the SMC may be understood
if the interstellar dust mixture includes 
magnetic nanoparticles, emitting magnetic dipole radiation resulting
from thermal fluctuations in the magnetization.
The magnetic grains can be metallic iron, magnetite Fe$_3$O$_4$, or 
maghemite $\gamma$-Fe$_2$O$_3$.
The required mass of iron is consistent with elemental abundance
constraints.
The magnetic dipole emission is predicted to be polarized orthogonally
to the normal electric dipole radiation if the nanoparticles are inclusions
in larger grains.
We speculate that other low-metallicity galaxies may also have a large fraction
of the interstellar Fe in magnetic materials.
\end{abstract}


\section{Introduction
         \label{sec:intro}}

Low-metallicity dwarf galaxies often exhibit
surprisingly strong emission at submillimeter 
and mm wavelengths
\citep[e.g.,][]{Galliano+Madden+Jones+etal_2003,
       Galliano+Madden+Jones+etal_2005,
       Galametz+Madden+Galliano+etal_2009,
       Grossi+Hunt+Madden+etal_2010,
       OHalloran+Galametz+Madden+etal_2010,
       Galametz+Madden+Galliano+etal_2011},
substantially exceeding what is expected
based on the observed emission from dust at shorter wavelengths.
This ``submm excess'' could in principle be due to
a large mass of cold dust,
but in some cases the implied
dust masses are too large
to be consistent with the observed gas mass 
and metallicity.

The Small Magellanic Cloud (SMC)
is a prime example of this phenomenon.
The dust spectral energy distribution (SED) 
has been measured from near-infrared through
cm wavelengths.
Both \citet{Bot+Ysard+Paradis+etal_2010} and
\citet{Planck_LMC_SMC_2011} conclude that
conventional dust models cannot account for the observed
3\,mm -- $600\micron$ (100\,GHz -- 500\,GHz)
emission without invoking unphysically large amounts
of very cold dust.

Large submm excesses have also been reported for other low-metallicity
dwarf galaxies.
NGC~1705 has received particular attention
\citep{Galametz+Madden+Galliano+etal_2009,
       OHalloran+Galametz+Madden+etal_2010,
       Galametz+Madden+Galliano+etal_2011}
and substantial excesses have also been reported for a number of
other systems, including
Haro~11, II~Zw~40, and NGC~7674
\citep{Galametz+Madden+Galliano+etal_2011}.

This excess emission 
challenges our understanding of interstellar dust.  If the
submm excess in low-metallicity dwarfs is due to thermal emission from
dust, these galaxies either contain surprisingly large masses of very cold
dust, or the dust opacity at submm frequencies must substantially exceed
that of the dust in normal-metallicity galaxies, such as the Milky Way.

In the Galactic ISM, typically
90\% or more of the Fe is missing from the gas phase
\citep{Jenkins_2009}, locked up in solid
grains.
Thus Fe accounts for $\sim$$25\%$ of the dust mass in diffuse
interstellar regions, 
although as yet we know little about the nature of the Fe-containing
material.
Interstellar dust models based on amorphous silicate and
carbonaceous material 
\citep[e.g.,][]{Mathis+Rumpl+Nordsieck_1977,
Draine+Lee_1984,
Weingartner+Draine_2001a,
Zubko+Dwek+Arendt_2004,
Draine+Li_2007,
Draine+Fraisse_2009}
often posit that the Fe missing from the gas
is incorporated in amorphous
silicate material, but it is entirely possible for much or most of the
solid-phase Fe to be in the form of metallic Fe or certain Fe oxides,
such as magnetite, that are spontaneously magnetized. 

\citet[][hereafter DL99]{Draine+Lazarian_1999a} noted that
ferromagnetic or ferrimagnetic materials can have large opacities
at microwave frequencies.
\citet{Draine+Hensley_2012a} recently re-estimated the absorption cross sections
for nanoparticles of ferromagnetic or ferrimagnetic materials.
They considered three 
naturally-occuring magnetic materials --
metallic iron, magnetite, 
and maghemite --
and found that the magnetic
response implies a large opacity
at submm and mm wavelengths.
 
In this {\it Letter} we propose that magnetic nanoparticles may
provide the 100--500$\GHz$ opacity needed to account for the
strong submm--microwave emission from the SMC.
Upper limits on dust masses in the SMC are
obtained in Section \ref{sec:smc_mass}.
The observed SED of the SMC, and the emission attributed to dust, 
is reviewed in Section
\ref{sec:smc spectrum}, and in Section \ref{sec:dl07 models} we
show that models with Milky Way
dust opacities cannot reproduce the observed SED.
The contribution of spinning dust is discussed in 
Section \ref{sec:spinning_dust}, where we show that spinning dust cannot
account for the observed emission near $\sim100\GHz$.
In Section \ref{sec:magnetic_dust_models} we consider dust models
for the SMC  
that include maghemite, magnetite, and metallic iron grains.
We find that the submm
and mm excess in the SMC can be accounted for by a population of
magnetic nanoparticles.

In Section \ref{sec:discussion} we discuss
other evidence for the formation of Fe or Fe-oxide nanoparticles,
and speculate on why the dust in low-metallicity galaxies such as the SMC
differs from the dust in normal metallicity spirals, such as the Galaxy. 
We also discuss the predicted polarization of the emission from
the SMC.
Our results are summarized in Section \ref{sec:summary}.

\section{\label{sec:smc_mass}
         Mass of the ISM in the SMC}

At a distance $D=62\kpc$ \citep{Szewczyk+Pietrzynski+Gieren+etal_2009}, 
the Small Magellanic Cloud (SMC)
provides an opportunity to study the dust in a low-metallicity dwarf galaxy.
The present study will concentrate on the $2.38^\circ$ 
radius ($\Omega=0.00542\sr = 17.8{\,\rm deg}^2$) region
(centered on $\alpha_{2000}=00^{\rm h}53^{\rm m}59.6^{\rm s}$,
$\delta_{\rm 2000}=-72^\circ40^\prime16.1^{\prime\prime}$) 
studied by \citet{Planck_LMC_SMC_2011}.
The 21$\cm$ line flux from this region at SMC radial velocities is 
$2100\Jy\,{\rm MHz}$ (mean line intensity $1318\K\kms$ over the aperture;
J.-P. Bernard 2012, private communication) corresponding to
optically-thin emission from $M({\rm H\,I})=3.99\times10^{8}\Msol$.
\citet{Stanimirovic+Stavely-Smith+Dickey+etal_1999} estimated that
correction for self-absorption would raise the \ion{H}{1} mass by
$0.42\times10^8\Msol$ (for $D=62\kpc$).
Thus we estimate $M({\rm H\,I})=4.41\times10^8\Msol$ 
in the $0.00542\sr$ region.
\citet{Leroy+Bolatto+Stanimirovic+etal_2007} find 
$M(\HH)\approx0.32\times10^8\Msol$.
Thus we take $M_\Ha \approx 4.73\times10^8\Msol$ (not including He) within the 
$0.00542\sr$ aperture.

Elemental abundances in the SMC are uncertain.
\citet{Russell+Dopita_1992} estimated
${\rm (Fe/H)}_{\rm SMC}=0.25{\rm (Fe/H)}_\odot$, while
\citet{Kurt+Dufour_1998} estimated 
${\rm (O/H)}_{\rm SMC}=0.2{\rm (O/H)}_\odot$.
\citet{Rolleston+Venn+Tolstoy+Dufton_2003} measured the abundances in
a main-sequence B star, and found 
${\rm (O/H)}=0.6{\rm (O/H)}_\odot$, and 
${\rm (Fe/H)}=0.3{\rm (Fe/H)}_\odot$.
\citet{Lee+Rolleston+Dufton+Ryans_2005} measured abundances in 3 B-type
supergiants in the SMC wing, finding
${\rm (Mg/H)}\approx 0.1{\rm (Mg/H)}_\odot$,
${\rm (Si/H)}\approx 0.2{\rm (Si/H)}_\odot$.
Here we adopt an overall metallicity $Z_{\rm SMC}\approx 0.25 Z_\odot$.

An upper limit on the dust mass in the SMC
is obtained from the observed gas
mass combined with the estimated abundances of elements that could
form dust grains.
The gas in diffuse H\,I and $\HH$ in the local interstellar medium
is routinely strongly depleted in elements such as Fe and Si, with the
missing material presumed to be in the form of dust.
An inventory of the well-studied sightline toward $\zeta$Oph
allows one to estimate the dust/gas mass ratio based on the
amount of C, O, Mg, Si, Fe, and other elements that are missing from
the gas \citep[][Table 23.1]{Draine_2011a}.\footnote{The sightline toward $\zeta$Oph is ice-free.  We assume that ices are a
negligible fraction of the total dust mass in the SMC.}
If we assume the relative abundances of heavy elements in a galaxy
to be similar 
to solar composition \citep{Asplund+Grevesse+Sauval+Scott_2009}, then
\beqa
M_{\rm dust}/M_\Ha &\leq& 0.0091 (Z/Z_\odot)
\\
M_{\rm Fe,dust}/M_\Ha &\leq& 0.00196 (Z/Z_\odot)
~~~,
\eeqa
where $M_{\rm Fe,dust}$ is the mass of Fe contained in solid material,
including ferromagnesian silicates, Fe oxides, and metallic Fe.
The mass of Fe in magnetic materials obviously is limited by
$M_{\rm Fe,dust}$.
The resulting upper limits on $M_{\rm dust}$ and Fe in magnetic form
are given in the first line of Table \ref{tab:masses}.

\begin{table}[ht]
\newcommand \fnab  {a}
\newcommand \fnbb  {b}
\newcommand \fnca  {b}
\newcommand \fncb  {c}
\newcommand \fndb  {c}
\begin{center}
\caption{\label{tab:masses}
         Dust Masses in the SMC
        }
{\footnotesize
\begin{tabular}{lccl}
\hline
          & Total Dust & Magnetic Fe \\
          & ($10^5 \Msol$)    & ($10^5\Msol$) & comment\\
\hline
{\bf Abundance limit}                             & {\bf${\bf\leq}$10.7}~~ 
                                                  & {\bf$\leq$2.3}~~
     & \\
Model 1: DL07 dust, $U_{\rm min}\geq0.2$          & 13.~ & -- 
     & violates abundance limit; poor fit \\
Model 2: DL07 dust, $U_{\rm min}\geq0.5$          & ~9.7 & --
     & very poor fit \\
Model 3: DL07 dust + $40\K$ Fe                    & ~8.3 & 1.4
     & OK \\
Model 4: DL07 dust + $20\K$ Fe                    & 10.2 & 2.2
     & OK \\
Model 5: DL07 dust + $17\K$ $\gamma$-Fe$_2$O$_3$  & ~9.4 & 2.2
     & OK \\
Model 6: DL07 dust + $17\K$ Fe$_3$O$_4$           & ~7.2 & 2.2
     & OK \\ 
\hline
\multicolumn{3}{l}{$\,^a$ Fe mass in magnetic material}\\
\multicolumn{3}{l}{$\,^b$ for $M_\Ha=4.7\times10^8\Msol$ and $Z=0.25Z_\odot$}\\
\multicolumn{3}{l}{$\,^\fnab$ $T=40\K$ Fe particles}\\
\multicolumn{3}{l}{$\,^\fnbb$ DL07 dust with cold component}\\
\multicolumn{3}{l}{$\,^\fncb$ $T=40\K$ Fe particles}\\
\end{tabular}
}
\end{center}
\end{table}

\section{\label{sec:smc spectrum}
         SED of the SMC}

\begin{figure}[t]
\begin{center}
\includegraphics[angle=270,width=8.0cm]{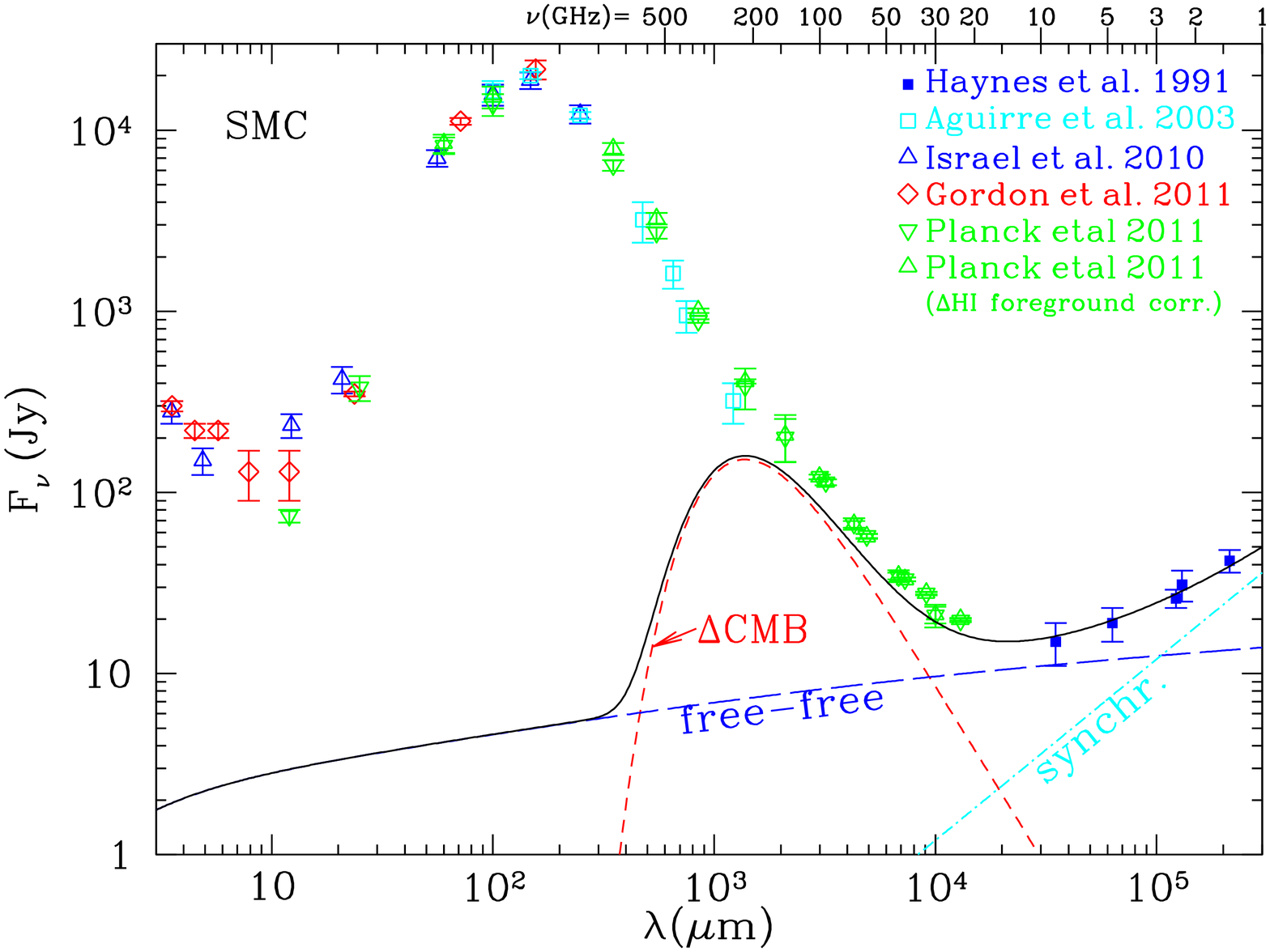}
\includegraphics[angle=270,width=8.0cm]{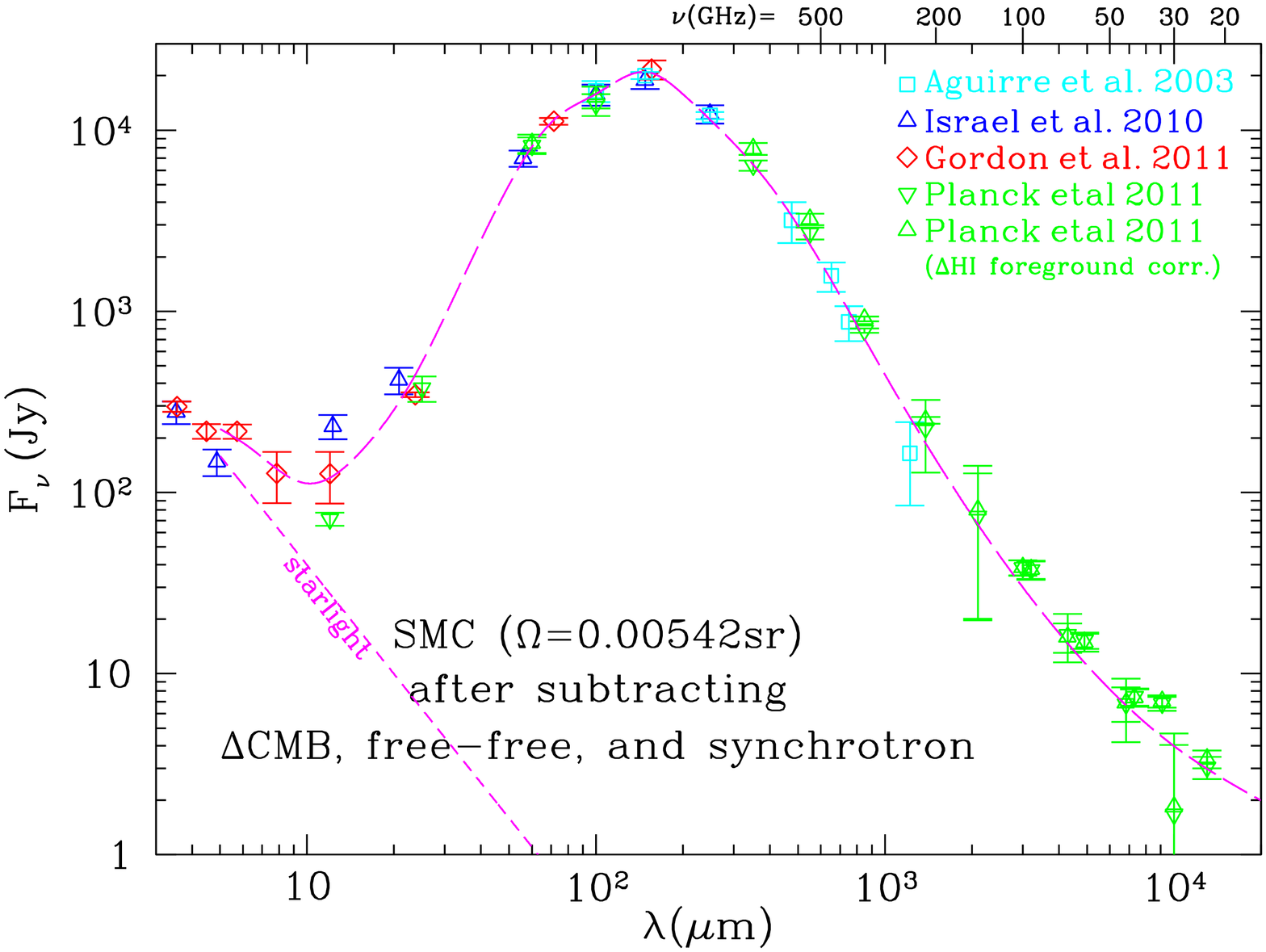}
\caption{\label{fig:SMC_SED}\footnotesize
   (a) The observed SED of the SMC (2.38$^\circ$ radius region),
   after subtraction of Galactic foreground emission.
   Estimated contributions from
   SMC synchrotron and free-free, and from CMB fluctuations
   are shown.
   (b) SED of the SMC after subtraction of $\Delta$CMB, free-free,
   and synchrotron.
  Flux densities measured by IRAS, COBE-DIRBE, TopHat, IRAC, MIPS, WMAP,
  and Planck are taken from tabulations by 
  \citet{Aguirre+Bezaire+Cheng+etal_2003},
  \citet{Israel+Wall+Raban+etal_2010},
  \citet{Gordon+Meixner+Meade+etal_2011},
  and
  \citet{Planck_LMC_SMC_2011} (see text).
  }
\end{center}
\end{figure}

Figure \ref{fig:SMC_SED} shows the observed global SED of the SMC, after
removal of smooth foregrounds and backgrounds,
from the following compilations:

\citet{Haynes+Klein+Wayte+etal_1991} reported global flux
densities measured with the Parkes 64m telescope
at 1.4, 2.45 GHz, 4.75, and 8.55GHz;
the 1.4 GHz flux density is a revision of the result of
\citet{Loiseau+Klein+Greybe+etal_1987}.
\citet{Mountfort+Jonas+deJager+Baart_1987} measured the 2.3 GHz flux
density with the Hartebeesthoeck 26m telescope.

The TopHat balloon experiment 
\citep{Aguirre+Bezaire+Cheng+etal_2003} measured the flux in 4 bands 
(245--630$\GHz$) in
a $2.40^\circ$ radius region ($\Omega=0.00544\sr$) centered on the SMC.
Foreground removal was done by subtracting the mean brightness of adjacent
off-source regions.
\citet{Aguirre+Bezaire+Cheng+etal_2003} 
also extracted 100, 140, and 240$\micron$ fluxes for
COBE-DIRBE \citep{Silverberg+Hauser+Boggess+etal_1993}.

\citet{Israel+Wall+Raban+etal_2010} extracted fluxes for a $2.40^\circ$
radius region ($\Omega=0.00544\sr$) centered on the SMC.
We show their extractions 
for the 10 COBE-DIRBE bands (1.27$\micron$ to 248$\micron$).

\citet{Gordon+Meixner+Meade+etal_2011} 
extracted fluxes for a $2.25^\circ$
radius region ($\Omega=0.00484\sr$) centered on the SMC,
measured using the 
IRAC \citep{Fazio+Hora+Allen+etal_2004} and 
MIPS \citep{Rieke+Young+Engelbracht+etal_2004} cameras
on the {\it Spitzer Space Telescope} \citep{Werner+Roellig+Low+etal_2004}.
Foreground removal consisted of subtracting the mean brightness of
an annulus extending from $2.3^\circ$ to $2.5^\circ$.

\citet{Planck_LMC_SMC_2011} extracted fluxes for a $2.38^\circ$ radius
region ($\Omega=0.00542\sr$) centered on the SMC.
Foreground subtraction consisted of subtracting the mean brightness
of a $1^\circ$ annulus around the extraction region.
We include their extractions for {\it Planck}
\citep[9 bands, 30--858 GHz][]{Planck_mission_2011},
{\it WMAP}
\citep[5 bands, 23--94 GHz][]{Bennett+Hill+Hinshaw+etal_2003}, 
and {\it IRAS} 
\citep[4 bands, 12--100$\micron$;][]{Miville-Deschenes+Lagache_2005}.
\citet{Planck_LMC_SMC_2011} further corrected the foreground removal
by taking into consideration the difference in $N($\ion{H}{1}$)$ at
Galactic radial velocities between the background annulus and the extraction
aperture.\footnote{
   We do not show the ``corrected'' Planck fluxes for IRAS12 and IRAS25
   because the entries for $I_\nu^{\rm sub}$ in Table 2 of
   \citet{Planck_LMC_SMC_2011} do not appear to be correct.
   }

Figure \ref{fig:SMC_SED}a shows the spectrum of the $0.00542\sr$ region
centered on the SMC.  
We assume that the differences in coverage ($\Omega$ ranging
from $0.00484\sr$ to $0.00544\sr$) are unimportant, as most of the
flux will come from the central regions.
\citet{Planck_LMC_SMC_2011} estimate that CMB fluctuations 
add emission corresponding to a mean
CMB temperature excess $\langle\Delta T_{\rm CMB}\rangle=58\mu{\rm K}$ over
the $0.00542\sr$ extraction region (relative to the background annulus).
The spectrum of this CMB excess 
\beq
(\Delta {\rm CMB})_\nu = \Omega \langle\Delta T_{\rm CMB}\rangle
\frac{\partial B_\nu}{\partial T} \bigg|_{T=2.726\K}
\eeq
is plotted in Figure \ref{fig:SMC_SED}a.

To isolate the emission from the dust, it is necessary to subtract
free-free and synchrotron emission.
We find the observations to be consistent with synchrotron and free-free
spectra
\beqa
F_\nu^{\rm synch} &\approx& 36. \left(\frac{\nu}{\GHz}\right)^{-1.0}\Jy
\\
F_\nu^{\rm ff} &\approx& 
11.0 \frac{g_{\rm ff}(\nu,T)}{g_{\rm ff}(10\GHz,T)}e^{-h(\nu-10\GHz)/kT} \Jy
\eeqa
with $T=10^4\K$.
Our estimate for $F_\nu^{\rm ff}(10\GHz)$ is
intermediate between the
$13.4\Jy$ estimate of \citet{Israel+Wall+Raban+etal_2010}
and the
$9.05\Jy$ estimate of \citet{Planck_LMC_SMC_2011}.
Our estimates for $F_\nu^{\rm ff}$ and $F_\nu^{\rm synch}$
are shown in Figure \ref{fig:SMC_SED}a.
The Gaunt factor $g_{\rm ff}(\nu,T)$ is obtained from eq.\ (10.9) of
\citet{Draine_2011a}.
For $n({\rm He}^+)/n({\rm H}^+)=1.08$ and $T=10^4\K$, 
this corresponds to
$\int n_e n({\rm H}^+)dV = 1.45\times10^{64}\cm^{-3}$
and an H photoionization rate
$Q_0=3.7\times10^{51}\s^{-1}$.

The residual after subtraction of 
$(\Delta{\rm CMB})_\nu$, $F_\nu^{\rm ff}$, and $F_\nu^{\rm synch}$
is shown in Figure \ref{fig:SMC_SED}b.  
This residual is presumed to
be emission from dust and (at short wavelengths) stars.
A smooth curve has been drawn through the points to guide the eye.
Subtracting an estimate for the starlight continuum as in Figure 1b, the
integrated $\lambda > 5\micron$ dust luminosity of the SMC is
$L_d(\lambda>5\micron)=1.00\times10^8(D/62\kpc)^2\Lsol$.

\section{\label{sec:dl07 models}
         Conventional Dust Models}

The observed infrared and submm emission from 
normal-metallicity star-forming spiral galaxies
appears to be consistent with physical dust models that were developed
to reproduce the observed properties of dust in the diffuse ISM of the local
Milky Way, including wavelength-dependent extinction
and infrared emission
\citep[e.g.,][]{Weingartner+Draine_2001a,
Li+Draine_2001b,
Zubko+Dwek+Arendt_2004,
Draine+Li_2007,
Draine+Fraisse_2009,
Compiegne+Verstraete+Jones+etal_2010}. 
The models of \citet[][henceforth DL07]{Draine+Li_2007}
consist of amorphous silicate grains plus carbonaceous grains;
the carbonaceous grains have
the physical properties of graphite when large, and the properties
of polycyclic aromatic hydrocarbons (PAHs) when very small.
This dust model was able to reproduce the global SEDs of the
galaxies in the SINGS sample \citep{Draine+Dale+Bendo+etal_2007}, including
17 galaxies with 850$\micron$ SCUBA photometry.
More recently, the same model has been found to be consistent with
both global and spatially-resolved SEDs of 
normal-metallicity (i.e., $0.5 \ltsim Z/Z_\odot \ltsim 2$) galaxies in the
KINGFISH sample
\citep{Aniano+Draine+Calzetti+etal_2012a,Aniano+Draine+Calzetti+etal_2012b},
including photometry out to 500$\micron$.

The DL07 dust model 
is able to reproduce the observed emission 
from dust in the diffuse ISM of the Galaxy 
\citep{Finkbeiner+Davis+Schlegel_1999}
out to wavelengths as long as $2\mm$ 
\citep[see Fig.\ 14a of][]{Draine+Li_2007}.
However, a significant emission excess (relative to the model) 
appears at $\lambda > 3\mm$ ($\nu<100\GHz$);
this ``anomalous microwave emission'' (AME) 
has been confirmed by numerous observations
\citep[][and references therein]{Planck_AME_2011}.
The AME in the Galaxy is thought to be
mainly rotational emission from the PAH population
\citep{Draine+Lazarian_1998a,Draine+Lazarian_1998b}.

DL07 propose that the $3\micron<\lambda<3\mm$ SEDs 
of entire galaxies, or large regions within a
galaxy, can be fit using a dust model consisting of amorphous
silicates, graphitic grains, and PAHs, and assuming that the dust
heating rate is distributed according to 
\beq \label{eq:dMd/dU}
\frac{dM_d}{dU} =
(1-\gamma)M_{d,tot} \delta(U-\Umin) + \gamma
M_{d,tot}\frac{(\alpha-1)U^{-\alpha}}{\Umin^{1-\alpha}-\Umax^{1-\alpha}}
~~~{\rm for}~~~\Umin\leq U\leq\Umax~~, 
\eeq where $U$ is the ratio of the local
dust heating rate to the heating rate produced by the
solar-neighborhood starlight radiation field, $M_d(U)$ is the mass
of amorphous silicate plus carbonaceous dust with heating rates $<U$,
and $M_{d,tot}$ is the total dust mass.  Eq.\ (\ref{eq:dMd/dU}) is a very simple
distribution function, with only 5 parameters ($M_{d,tot}$, $U_{\rm min}$, 
$U_{\rm max}$, $\alpha$, $\gamma$), but studies of dust emission
using this distribution function for the grain heating have been
successful in reproducing the global emission from
galaxies, as well as the emission from $\sim$500~pc regions within
galaxies \citep[e.g.,][]{Aniano+Draine+Calzetti+etal_2012a}.  The DL07
models are also characterized by the PAH abundance parameter 
$q_{\rm PAH}=$ the fraction of the dust mass contributed by
PAH particles containing $< 10^3$ C atoms.

\begin{figure}[t]
\begin{center}
\includegraphics[angle=270,width=8.0cm]{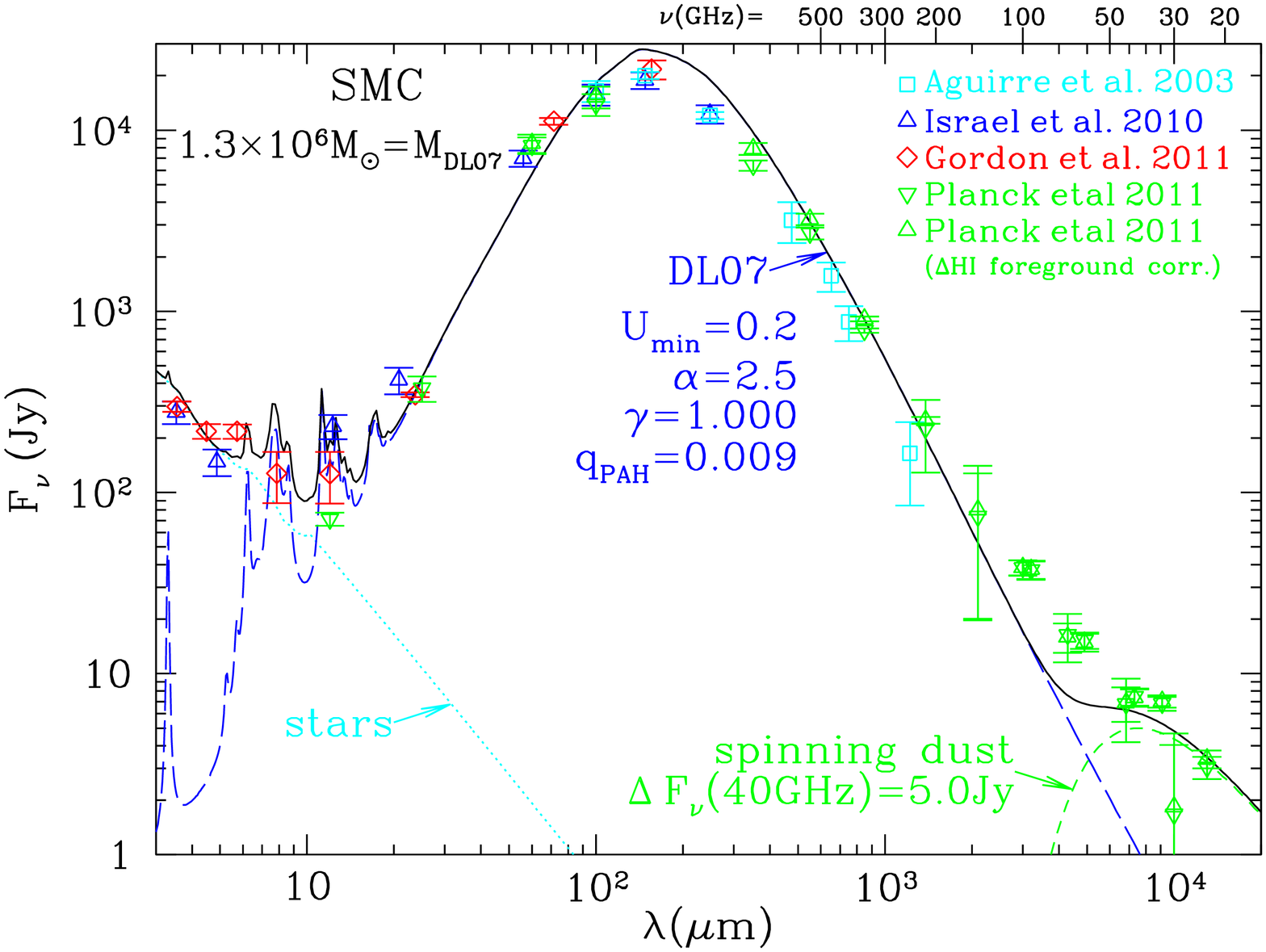}
\includegraphics[angle=270,width=8.0cm]{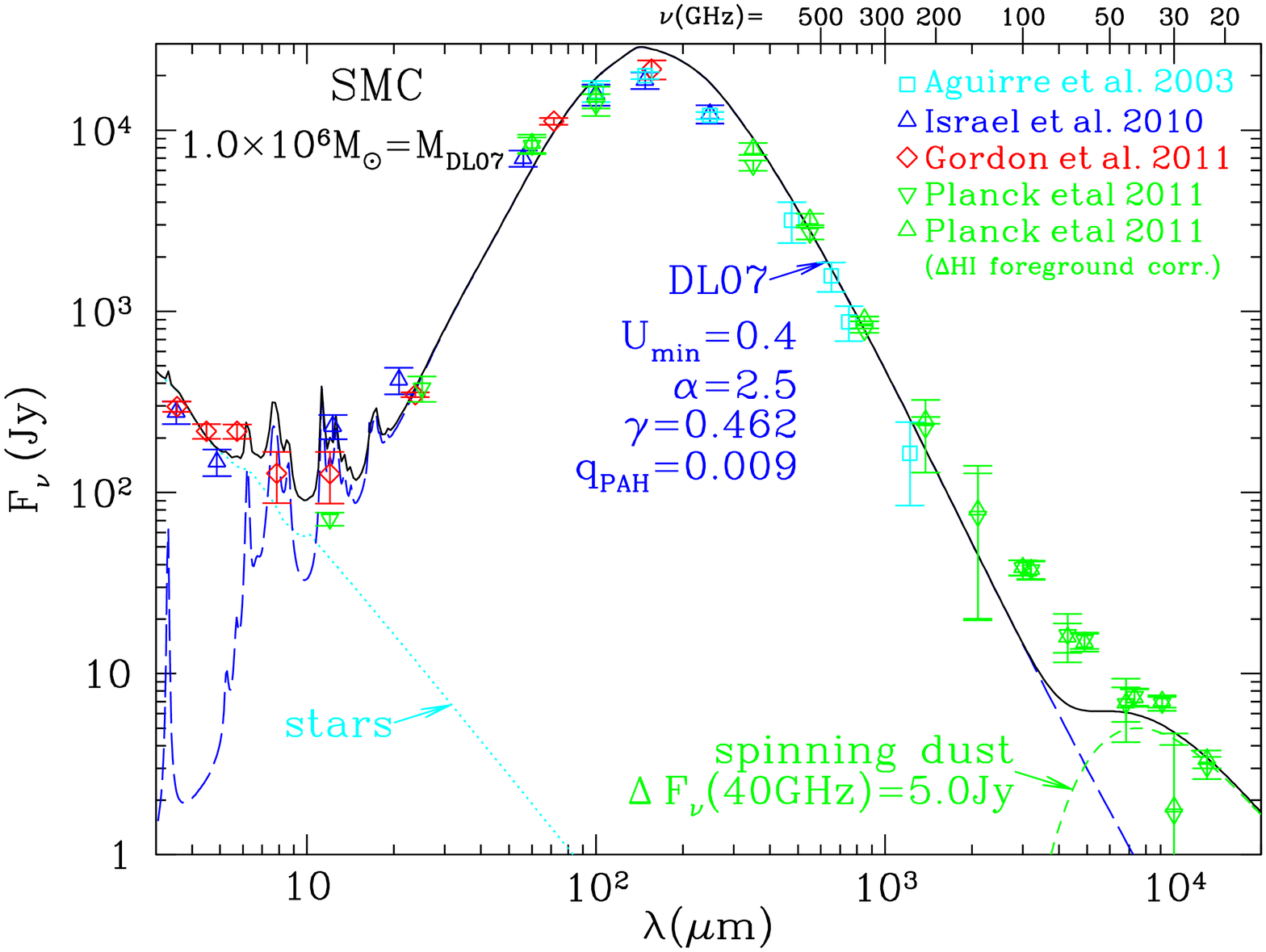}
\caption{\label{fig:SMC_SED_DL07}\footnotesize
  Data: Observed SED of the SMC, after removal of
  CMB fluctuations, and subtraction of free-free and synchrotron
  emission (see Figure \ref{fig:SMC_SED}).
  Solid line: model consisting of the sum of $T=5000\K$ starlight
  (dotted line),
  emission from the DL07 dust model (dot-dash line),
  and a spinning dust component peaking at $40\GHz$ (dashed line).
  (a) The total dust mass exceeds the maximum allowed
      by a factor $\sim$$20\%$ (see Table \ref{tab:masses}).
  (b) A model with the dust mass within allowed limits.
  In both models, the spinning dust component has been
  adjusted to reproduce the observed 20--50$\GHz$ emission.
  Both models provide
  insufficient 60--300$\GHz$ emission.
  The models in (a) and (b) have dust luminosities $L_d=7.0\times10^8\Lsol$
  and $6.9\times10^8\Lsol$.  
  }
\end{center}
\end{figure}

We vary 5 parameters -- the total dust mass $M_{d,tot}$,
the PAH abundance parameter $q_{\rm PAH}$,
and the starlight heating parameters
$\Umin$, $\alpha$, and $\gamma$; 
$\Umax=10^7$ is kept fixed.
Because we do not include a realistic model for the starlight contribution to
the SED, the
model is fit only to data at $\lambda > 3\micron$, where reddening by
dust should be minimal.
Because the emission at $\lambda\gtsim3\mm$ ($\nu\ltsim100\GHz$)
may include a substantial contribution from ``spinning dust'',
the DL07 model + starlight is fit only to $\lambda < 2\mm$ data.

If we allow $\Umin$ to be as low as $0.2$, we obtain Model 1,
shown in Figure \ref{fig:SMC_SED_DL07}.
This model has a total dust mass $M_{d,tot}=1.3\times10^6\Msol$, exceeding
the upper limit of $1.1\times10^6\Msol$
(see Table \ref{tab:masses}).  Despite using more dust than is allowed,
Model 1 provides insufficient emission at $\lambda>2\mm$.

Because Model 1 violates the dust abundance limit, we try fitting the
DL07 model to the same data, but now limiting $\Umin\geq 0.4$.
The resulting Model 2 has a total dust mass that does not violate the
upper limit in Table \ref{tab:masses}, but
the quality of the fit to the SED is somewhat
worse than for Model 1, with an
even larger deficiency at $\lambda > 2\mm$ (see Fig.\ \ref{fig:SMC_SED_DL07}b).

\section{\label{sec:spinning_dust}
         Spinning Dust}

We can add a spinning dust component to raise the emission in the
20--60$\GHz$ range.
The spinning dust emission 
in the diffuse ISM of the Galaxy peaks
at about 40 GHz; we expect a similar peak frequency in the SMC.
What level of emission is expected for spinning dust in the SMC?

\citet{Draine+Lazarian_1998a,Draine+Lazarian_1998b} argued
that the anomalous microwave emission in the Galaxy,
with an observed emissivity per H nucleon
$\left[j_\nu^{\rm (sd)}(40\GHz)/\nH\right]_{\rm MW}
\approx1\times10^{-17}\Jy\sr^{-1}\cm^2\Ha^{-1}$,
is primarily rotational emisssion from the PAH population.
The PAH abundance is measured by $\qpah$.
Dust in the solar
neighborhood is thought to have $q_{\rm PAH}\approx 4.7\%$.
\citet{Li+Draine_2002c} found that $q_{\rm PAH}$ in the SMC
was spatially variable and, on average,
much lower than in the Milky Way.
\citet{Sandstrom+Bolatto+Draine+etal_2010} confirmed this, finding
a mean $\langle q_{\rm PAH}\rangle \approx 0.6\%$.
We expect the dust/gas ratio in the SMC to be lower
by about a factor $\sim$$Z_{\rm SMC}/Z_\odot\approx0.25$.
Therefore the PAH abundance per H is down by about
a factor $\sim$$(0.6/4.6)\times0.25=0.033$.
Thus we
estimate the spinning dust emission
in the SMC to be
\beqa
\left[\frac{j_\nu^{\rm (sd)}(40\GHz)}{\nH}\right]_{\rm SMC}
&\approx& 0.033\times1\times10^{-17}\Jy\sr^{-1}\cm^2\Ha^{-1} \approx 3.3\times10^{-19}\Jy\sr^{-1}\cm^2\Ha^{-1}
\\ \label{eq:F_nu(sd)}
\Delta F_\nu^{\rm (sd)}(40\GHz)&\approx&
\left[\frac{j_\nu^{\rm (sd)}(40\GHz)}{\nH}\right]_{\rm SMC}\times \frac{M_\Ha}{m_\Ha}D^{-2} \approx 5\Jy
~~~.
\eeqa
In Figure \ref{fig:SMC_SED_DL07} we have added an emission component
with a spectrum\footnote{This simple form,
   adequate for the present purposes,
   approximates the spectra obtained by detailed calculations
   \citep[e.g.,][]{Draine+Lazarian_1998b,
                   Ali-Haimoud+Hirata+Dickinson_2009,
                   Hoang+Draine+Lazarian_2010,
                   Hoang+Lazarian+Draine_2011,
                   Silsbee+Ali-Haimoud+Hirata_2011,
                   Ysard+Juvela+Verstraete_2011}.
   }
\beq
\Delta F_\nu^{\rm (sd)} = \Delta F_\nu^{\rm (sd)}(\nu_0) 
\left(\frac{\nu}{\nu_0}\right)^2
\exp\left[1-(\nu/\nu_0)^2\right]
\eeq
representative of what is expected
for spinning dust.
If we set
$\nu_0=40\GHz$ and $\Delta F_\nu^{\rm (sd)}(40\GHz)=5\Jy$
-- consistent with
the estimate in Eq.\ (\ref{eq:F_nu(sd)}) -- the 
20--50$\GHz$ observations are accounted for, as seen in
Figure \ref{fig:SMC_SED_DL07}.

While spinning dust appears able to account for
the observed 20--50$\GHz$ emission, the emission between 
50 and 300$\GHz$ remains much stronger than expected.
\citet{Bot+Ysard+Paradis+etal_2010} suggest that the 50--300$\GHz$
excess could also be due to spinning dust emission.
However, theoretical models of
rotational emission from small grains
\citep{Draine+Lazarian_1998b,
Ali-Haimoud+Hirata+Dickinson_2009,
Hoang+Draine+Lazarian_2010,
Hoang+Lazarian+Draine_2011,
Silsbee+Ali-Haimoud+Hirata_2011}
predict little rotational emission above
$\sim$$100\GHz$
unless high densities and warm gas temperatures
are present in the emitting regions.
For example, \citet[][see Fig.\ 13]{Draine+Lazarian_1998b} 
calculated the spinning dust emission from a model PDR
with $\nH=10^5\cm^{-3}$,
$T=300\K$, illuminated by a radiation field $U\approx 3000$.
The PDR was assumed to have abundances of small grains relative
to big grains reduced by a factor of 5 relative to
diffuse clouds in the solar neighborhood, approximating the
observed reduction in $q_{\rm PAH}$ in the SMC.
Viewed face-on, the total IR luminosity/area
$L_{\rm TIR}/A
\approx3.6\times10^{-3}U \erg\cm^{-2}\s^{-1} = 11 \erg \cm^{-2}\s^{-1}$.
The spinning dust emission for this model peaked near $110\GHz$, with
$(\nu L_\nu^{\rm (sd)})_{100\GHz}/A= 7.4\times10^{-7}\erg\cm^{-2}\s^{-1}$.
Thus $L_{\rm TIR}/(\nu L_\nu^{\rm (sd)})_{100\GHz} \approx 1.5\times10^{7}$.

At $100\GHz$, the model in Figure \ref{fig:SMC_SED_DL07}b
has a deficit  $\Delta F_\nu\approx25\Jy$, corresponding to
$(\nu \Delta L_\nu)_{100\GHz}=4\pi D^2 (\nu \Delta F_\nu)_{100\GHz}= 3000\Lsol$.
To account for this would require PDRs with a luminosity
$L_{\rm IR}=4.5\times10^{10}\Lsol$ -- completely inconsistent with the
observed $L_{\rm TIR}=1\times10^8\Lsol$.
It is evident that spinning dust cannot account for the observed
50--200 GHz emission in the SMC.
Here we consider magnetic dust grains as an alternative.

\section{\label{sec:magnetic_dust_models}
         SMC Dust Models Including Magnetic Dust}

\begin{figure}[t]
\begin{center}
\includegraphics[angle=270,width=8.1cm]{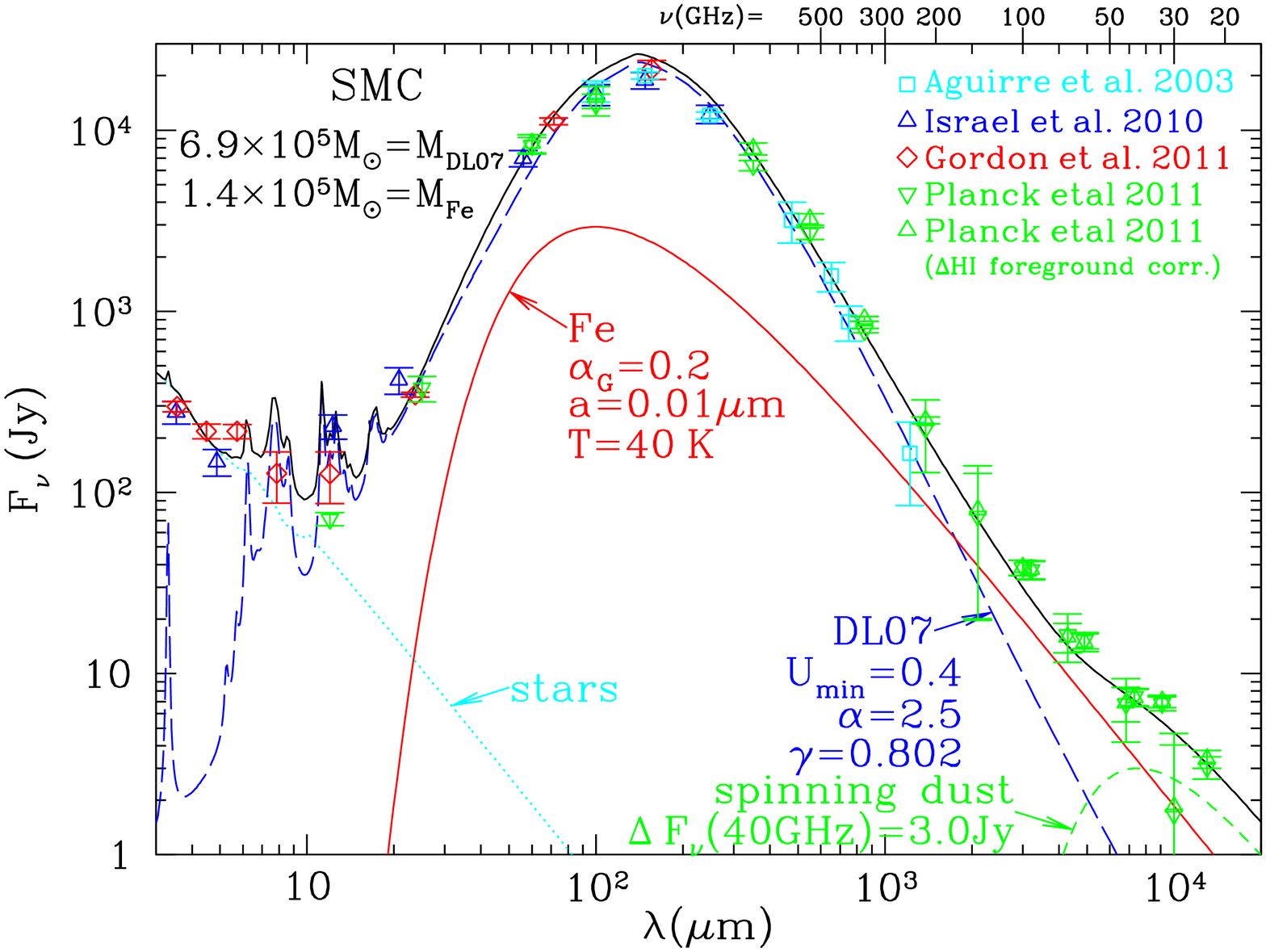}
\includegraphics[angle=270,width=8.1cm]{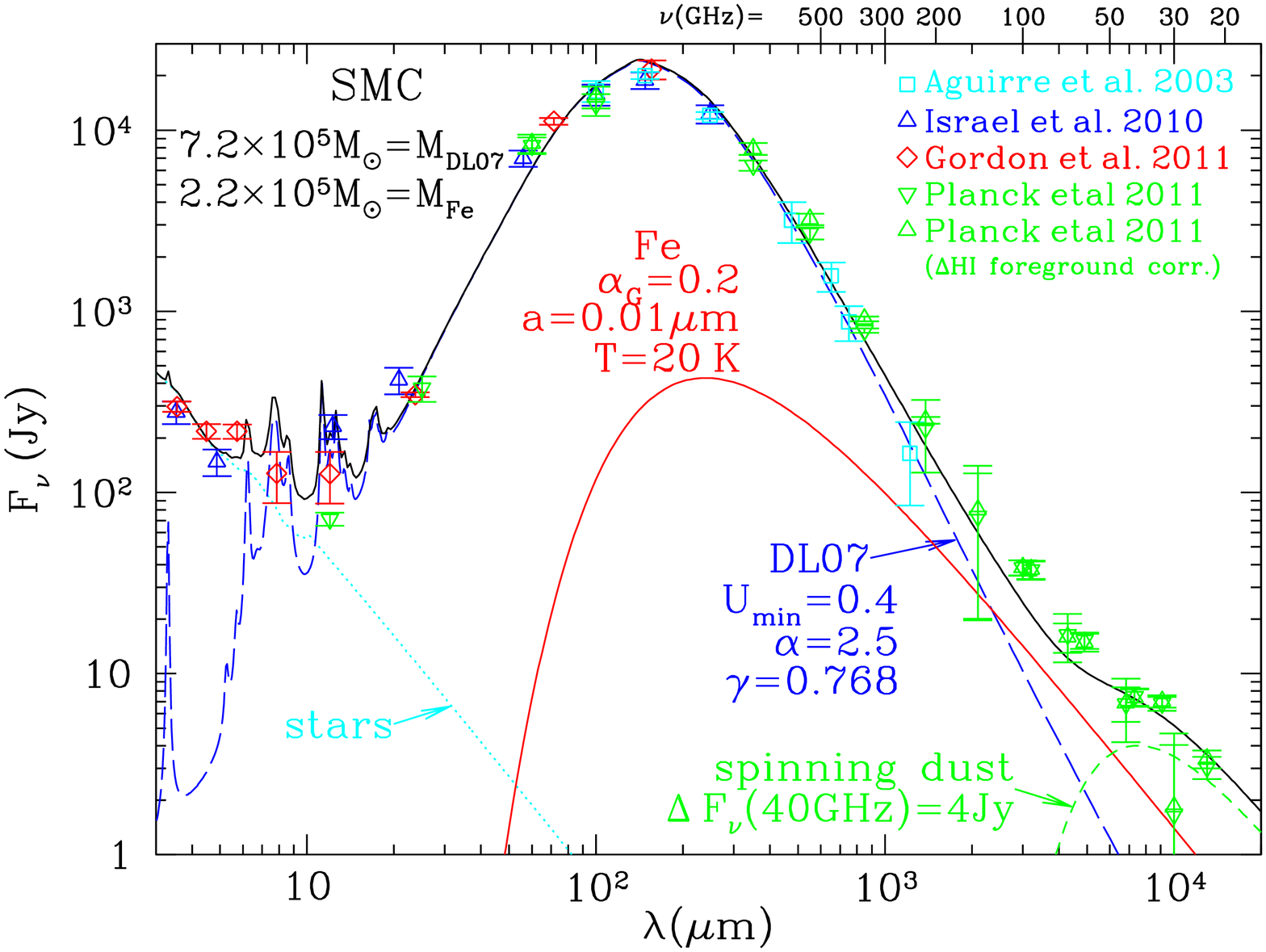}
\caption{\label{fig:SMC_SED_Fe}\footnotesize
         Similar to Figure \ref{fig:SMC_SED_DL07}, but with
         metallic Fe nanoparticles added to the dust model.
         The
         Fe particles are assumed to be at $T=40\K$ in Model 3 (panel a)
         and $T=20\K$ in Model 4 (panel b).
         }
\end{center}
\end{figure}

\begin{figure}[t]
\begin{center}
\includegraphics[angle=270,width=8.1cm]{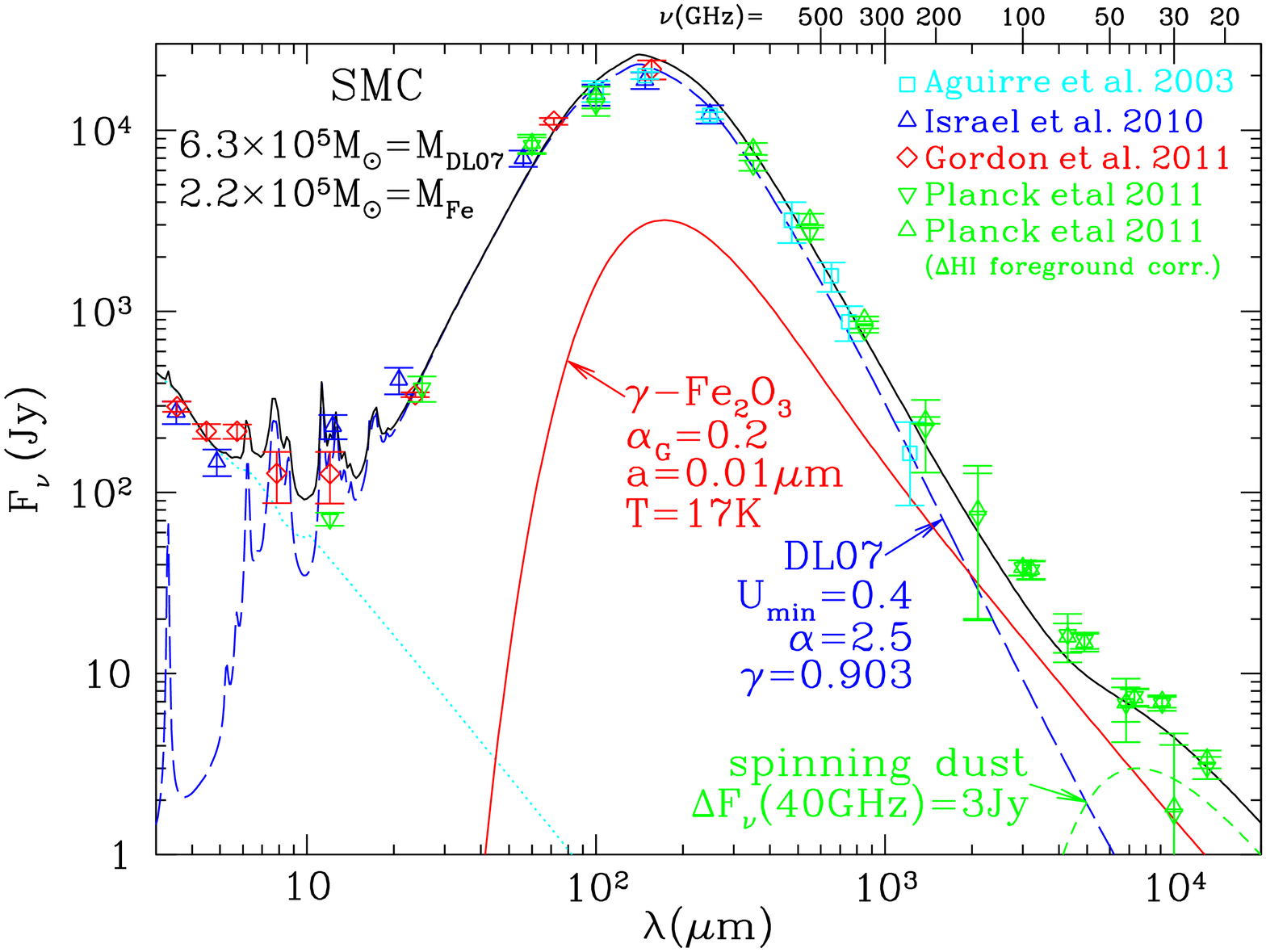}
\includegraphics[angle=270,width=8.1cm]{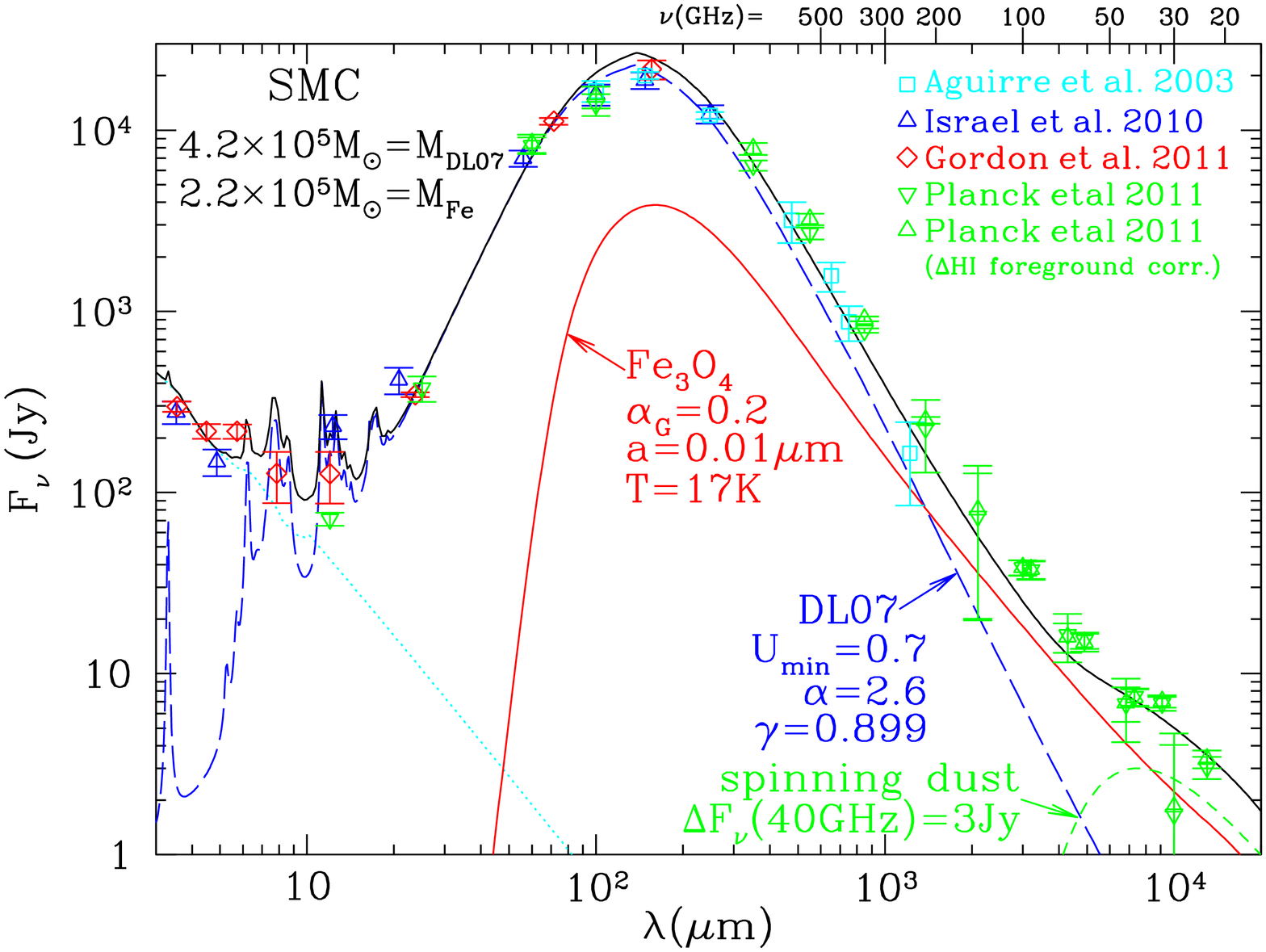}
\caption{\label{fig:SMC_SED_Feoxides}\footnotesize
         Similar to Fig.\ \ref{fig:SMC_SED_Fe}
         but with emission from (a) maghemite (Model 5) and (b)
         magnetite (Model 6) nanoparticles at $T=17\K$.
         }
\end{center}
\end{figure}
Because magnetic materials have enhanced absorption at 
microwave and submm
frequencies, it is of interest to see whether the mm- and cm-excess
seen in the SMC could be due in part to thermal emission from
magnetic grain materials.
In Figures \ref{fig:SMC_SED_Fe} and \ref{fig:SMC_SED_Feoxides}
we model the observed emission from the
SMC as the
sum of three components:  
``normal'' dust (the amorphous silicate, graphite, and PAH model
of DL07), a population of magnetic grains, and spinning dust.
In each case, the spinning dust contribution is assumed to peak at 40$\GHz$,
with the peak flux density adjusted to fit the observations
in Figure \ref{fig:SMC_SED}a, giving a reasonably good fit
in the $20-60\GHz$ region. 

Metallic iron nanoparticles are introduced in Fig.\ \ref{fig:SMC_SED_Fe}.
We consider Fe grain temperatures of 40$\K$ (Fig.\ \ref{fig:SMC_SED_Fe}a)
and 20$\K$ (Fig.\ \ref{fig:SMC_SED_Fe}b). 
If the Fe nanoparticles are, for the most
part, free-fliers heated by typical starlight, 
then $T\approx40\K$ is 
expected \citep[see Fig.\ 4 of][]{Draine+Hensley_2012a}.
If, on the other hand, 
the Fe nanoparticles are inclusions in larger composite grains, then
the $T\approx20\K$ temperature is appropriate, 
consistent with the temperature of
the ``normal'' dust.  
In each case, the Fe grain abundance is adjusted
to reproduce most of the observed emission near $100\GHz$, then 
a model using DL07 dust is used
to provide the additional emission required to reproduce
the observed SED at shorter wavelengths, and
finally a spinning dust component peaking at 40~GHz is added to bring the model
into agreement with the 20--50~GHz observations.

We also consider nanoparticles of maghemite
(Fig.\ \ref{fig:SMC_SED_Feoxides}a) and
magnetite (Fig.\ \ref{fig:SMC_SED_Feoxides}b).  For these we assumed
temperatures $T\approx 17\K$
consistent with being inclusions within nonmagnetic dust grains.

The model with maghemite (Fig.\ \ref{fig:SMC_SED_Feoxides}a)
has
$M_{\rm Fe}=2.2\times10^5\Msol$ of Fe in maghemite
(total maghemite mass $3.1\times10^5\Msol$)
and the model with magnetite (Fig.\ \ref{fig:SMC_SED_Feoxides}b) 
has
$M_{\rm Fe}=2.2\times10^5\Msol$ of Fe in magnetite
(total magnetite mass $3.0\times10^5\Msol$).
The Fe mass, and total dust mass, does not violate the mass
budget (see Table \ref{tab:masses}).
We conclude that the observed mm-wave emission from the SMC
can be accounted for by models with reasonable abundances of
normal dust plus 
metallic Fe, maghemite, magnetite, or some combination
of these three materials.

\begin{figure}[t]
\begin{center}
\includegraphics[angle=270,width=8.1cm]{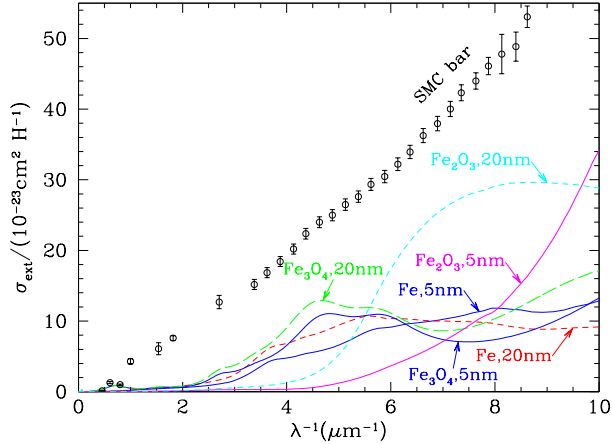}
\caption{\label{fig:SMC_ext}\footnotesize
         Extinction contributed by nanoparticles if 100\% of the Fe is
         in particles of 
         Fe, Fe$_3$O$_4$, or
         $\gamma$-Fe$_2$O$_3$, and radii $a=5\nm$ or
         $a=20\nm$.
         Also shown (symbols) is the observed extinction in the 
         SMC Bar
         \citep{Gordon+Clayton+Misselt+etal_2003}.
         The observed extinction does not
         rule out the hypothesis that most of the Fe is
         in free-flying nanoparticles.
         }
\end{center}
\end{figure}
If the nanoparticles are present as inclusions in larger grains, it is
clear that the size distribution of the larger particles can be adjusted
to be compatible with the observed wavelength-dependent extinction in the
SMC.
But is it possible for the bulk of the interstellar Fe to be in free-flying
nanoparticles?
We have calculated the extinction contribution in the
optical and UV, assuming that 100\% of the Fe is in particles of
a single type, and using dielectric functions for Fe, Fe$_3$O$_4$,
and $\gamma$-Fe$_2$O$_3$ from
\citet{Draine+Hensley_2012a}.
Figure \ref{fig:SMC_ext} shows the calculated extinction per H,
together with the observed extinction
in the SMC Bar \citep{Gordon+Clayton+Misselt+etal_2003}.
In no case does the calculated extinction/H exceed the observed extinction.
Therefore, the observed extinction is not incompatible with the possibility
that much of the Fe is in free-flying nanoparticles.

Nevertheless, we consider it most 
likely that the bulk of the magnetic nanoparticles would
be present as inclusions in larger grains, since we know that
most of the grain mass is in grains with radii $a\gtsim0.1\micron$

\section{\label{sec:discussion}
         Discussion}

\subsection{Solid-Phase Iron in Low Metallicity Galaxies}

At submm - mm frequencies, the
SED of the SMC is significantly less steep than the
SED of normal-metallicity spiral galaxies, including the Galaxy.
If the enhanced emission of the SMC at mm-wavelengths is due to
Fe or Fe oxide grains, then such grains must account for a larger fraction
of the dust mass in the SMC than in normal-metallicity spirals:
conditions in the SMC must be in some way more
favorable for their production or survival than in normal star-forming
galaxies. 



{\it Spitzer Space Telescope} observations of globular clusters have
detected excess infrared emission from the most luminous giant stars,
indicative of dusty winds.
In many cases, the
IR spectrum of the infrared excess 
is dominated by a featureless
continuum at $\lambda > 8\micron$.
Globular clusters where such featureless spectra have been seen
include 
47 Tuc 
\citep{McDonald+Boyer+vanLoon+Zijlstra_2011},
NGC 362 \citep{Boyer+McDonald+vanLoon+etal_2009},
and
Omega Cen 
\citep{McDonald+vanLoon+Sloan+etal_2011}; these
three clusters have metallicities
[Fe/H]$\approx-0.7$,
$-1.3$,
and $-1.5$, respectively
\citep[][2010 edition]{Harris_1996}\footnote{
The updated Harris cluster catalogue can be found at
\url{http://www.physics.mcmaster.ca/$\sim$harris.mwgc.dat}}.
The featureless emission might be attributed to hot amorphous carbon grains,
but carbon solids are not expected to form in these oxygen-rich outflows.
The featureless continuum has therefore instead been
attributed to thermal emission from
metallic Fe grains
\citep{McDonald+Sloan+Zijlstra+etal_2010,
       McDonald+Boyer+vanLoon+Zijlstra_2011,
       McDonald+vanLoon+Sloan+etal_2011}.
Thus, low-metallicity AGB stars provide a possible source for metallic
iron or Fe oxide grains in low-metallicity galaxies such as the SMC. 

The iron-rich ejecta of
Type Ia supernovae
constitute a second potential source of iron grains
\citep{Dwek_1998}.
However, to date there is no evidence of dust formation in SN Ia ejecta,
despite sensitive searches
toward the Tycho and Kepler SNRs
\citep{Gomez+Clark+Nozawa+etal_2012}.

Type II supernovae are known to form dust in the ejecta in at least
some cases
\citep{Sugerman+Ercolano+Barlow+etal_2006,Matsuura+Dwek+Meixner+etal_2011},
and it is conceivable that Fe-rich portions of the ejecta might
condense metallic Fe or Fe oxides.
\citet{Rho+Kozasa+Reach+etal_2008} made models to reproduce the
5--38$\micron$ spectra of the Cas A ejecta; their global model had 
$0.028\Msol$ of dust, of which 37\% was metallic Fe.

\citet{Baron+Bilson+Gold+etal_1977} observed that lunar soil grains
have an increase in the concentration of Fe near the surface, with
some of the Fe in metallic form.
These surface layers (``rims'') reflect exposure of the grains
to cosmic rays and the solar wind.
``Inclusion-rich rims''
consist of an amorphous silica-rich matrix with abundant metallic Fe
inclusions, typically $<10\nm$ in diameter
\citep{Keller+McKay_1997}.
Inclusion-rich rims are compositionally distinct from the host grain, and
are thought to have formed by deposition of atoms from vapors
produced by nearby sputtering or impact events.
In the laboratory, 
irradiation of olivine by 4~keV He ions is observed to lead to alteration
of the surface layers, 
with formation of metallic Fe nanoparticles
\citep{Dukes+Baragiola+McFadden_1999,
       Carrez+Demyk+Cordier+etal_2002,
       Loeffler+Dukes+Baragiola_2009}.
Metallic Fe nanoparticles are found as inclusions 
in interplanetary dust particles known as
GEMS 
\citep[``Glasses with Embedded Metals and Sulfides'';][]{Bradley_1994}.
Thus, it is reasonable to consider that some 
of the Fe in interstellar grains may
be in metallic Fe inclusions.

Fe-rich grain material
is injected into the ISM from stellar sources; as seen
above, this may include metallic Fe.
Additional conversion of gas-phase
Fe to solids must take place in the ISM to account for observed low gas-phase
abundance of Fe, particularly in view of the likely importance
of grain destruction by sputtering in supernova blastwaves
\citep{Draine+Salpeter_1979b,
       Jones+Tielens+Hollenbach+McKee_1994,
       Draine_2009b}:
it has been estimated that
``stardust'' (material condensed in stellar outflows)
accounts for only a small fraction
-- perhaps 10\% -- of the interstellar grain mass in the Galaxy
\citep{Draine_1990,Draine_2009b}, and this is likely the case
for all galaxies where a substantial fraction of the
refractory elements (Mg, Si, Fe) is in grains.
In such galaxies, including the SMC,
the bulk of the grain material must have undergone
conversion from gas to solid in the ISM.
The character of the interstellar dust will therefore 
be largely determined by interstellar processing.

Sputtering by energetic H and He 
can alter the composition of interstellar dust.
Sputtering yields have been discussed by a number of authors
\citep[e.g.,][]{Draine+Salpeter_1979a,Tielens+McKee+Seab+Hollenbach_1994}.
For a composite material, sputtering yields for H and He will be larger
for the lighter elements in the target, and sputtering will therefore leave
the surface layers enriched in heavy elements
(such as Fe).  The grain material that survives sputtering will therefore
become Fe-rich, perhaps even metallic Fe.
Studies of elemental depletions in the solar neighborhood indeed
suggest that Fe is concentrated in grain cores
\citep{Fitzpatrick+Spitzer_1997,Jenkins_2009}.
Based on the observed depletion patter toward Sk~155 in the SMC,
\citet{Welty+Lauroesch+Blades+etal_2001} suggested that much of the
interstellar Fe in the SMC 
(at least on the sightline to Sk 155) 
is in the form of metallic Fe or Fe oxides.

Rates for grain growth by accretion are proportional to the metallicity,
while rates for grain destruction by H and He sputtering are not.
The balance between
grain growth and destruction,
and the composition of the extant material, will therefore depend on the
metallicity of the ISM.
Hence, the apparent difference in
grain composition between normal-metallicity spirals (like the Milky Way)
and low-metallicity dwarf galaxies such as the SMC.

\subsection{High-Frequency Magnetism and the Gilbert Equation}

The models presented here use absorption cross sections $C_\abs(\omega)$
for magnetic grains calculated following
\citet{Draine+Hensley_2012a}, who used
the Gilbert equation \citep{Gilbert_2004}
to model the frequency-dependent magnetic response of Fe, 
maghemite, and magnetite.
The Gilbert equation uses
an adjustable dimensionless parameter $\alphaG$ to characterize the dissipation.
We have adopted $\alphaG\approx0.2$ for purposes of discussion, 
but
the existing experimental literature employs a range of values of
$\alphaG$.
If $\alphaG$ were to be smaller than $0.2$,
the $\nu\gtsim 100\GHz$ opacity would be reduced,
and 
the mass of Fe required to reproduce the observed emission of the SMC
would correspondingly increase.
If $\alphaG \ltsim 0.05$, magnetic
grain models to explain the $\sim$$3\mm$ emission would be ruled out by abundance
constraints.

Quite aside from the question of what value to use for $\alphaG$, 
it is also important to recognize that the
prescription for dissipation in the Gilbert equation,
while mathematically convenient, is not based on an underlying
physical model.  Empirical evidence
for the accuracy of the Gilbert equation at high frequencies is scant.
Laboratory measurements of electromagnetic absorption
in Fe and Fe oxide nanoparticles at frequencies up to 500 GHz are
needed to validate use of the Gilbert equation at these frequencies.

\subsection{Polarization}

Based on starlight polarization studies, the magnetic field in the SMC
appears to lie primarily in the plane of the sky, with substantial
large-scale coherence
\citep{Mao+Gaensler+Stanimirovic+etal_2008,
       Mao+Gaensler+Stanimirovic+etal_2012}.
While our understanding of the physics of grain alignment remains incomplete,
dust grains in the SMC are expected to be partially
aligned with long axes tending to be perpendicular 
to the local magnetic field $\bB_0$.
Electric-dipole emission, which dominates for $\lambda \ltsim 500\micron$,
will be polarized with $\bE_\omega\perp\bB_0$.

As seen above,
magnetic dipole emission from magnetic nanoparticles
may become important for $\lambda \gtsim 1\mm$.
The polarization of the magnetic dipole emission has been discussed
by \citet{Draine+Hensley_2012a}.  
If the magnetic nanoparticles are
free-fliers, and are aligned by a Davis-Greenstein-like mechanism,
the magnetic dipole emission will be polarized in the same sense
as the FIR emission, but the fractional polarization may be even larger than
that of the FIR emission (see Fig.\ 9 of \citet{Draine+Hensley_2012a}).
Alternatively, 
if the magnetic nanoparticles are present as randomly-oriented
inclusions within larger aligned grains,
the magnetic dipole emission will be 
polarized with $\bE_\omega \parallel \bB_0$.
As a result
the fractional polarization may decrease by a factor $\sim$$2$ as the
frequency decreases from $200\GHz$ to $\sim$$40\GHz$,
and for magnetite, maghemite, or Fe spheroids, the net polarization
undergoes a reversal (i.e, changes from $\bE_\omega\perp\bB_0$ to 
$\bE_\omega\parallel\bB_0$) near $\sim$$15\GHz$
\citep[see Fig.\ 10 of][]{Draine+Hensley_2012a}.

{\it Planck} will measure the polarization at 
$30$, $44$, $70$, $143$, $217$, and $353\GHz$.
Unfortunately, there are two additional factors that will complicate
interpretation of the dependence of polarization fraction on
frequency:
\begin{itemize}
\item Emission from spinning dust
becomes increasingly important with decreasing frequency, 
peaking near $\sim$$40\GHz$.
If this emission component is
minimally polarized, as predicted \citep{Lazarian+Draine_2000},
it will cause the fractional
polarization to {\it decrease} 
with decreasing frequency in the 60--$100\GHz$ region.
\item
There may be more than one grain type contributing to the
normal ``electric dipole'' emission at long wavelengths, as in the
mixtures of silicate and carbonaceous grains considered by
\citet{Draine+Fraisse_2009}.  
In this case, even the ``normal'' electric dipole emission
alone may have the fractional polarization depending significantly
on frequency.
In the models of \citet{Draine+Fraisse_2009}, the fractional polarization
is predicted to {\it increase} with decreasing frequency.
\end{itemize}
Actual reversal of the polarization
below $\sim$$15\GHz$ would be an unambigous indication of
magnetic dipole emission from magnetic inclusions, 
but this may be overwhelmed
by the increasing importance of synchrotron emission
(polarized with $\bE_\omega\perp\bB_0$)
as the frequency falls below $10\GHz$.

\section{\label{sec:summary}
         Summary}

The principal conclusions of this paper are as follows:
\begin{enumerate}

\item We show (see Figs.\ \ref{fig:SMC_SED_Fe} and \ref{fig:SMC_SED_Feoxides})
that the SED of the SMC can be approximately reproduced
by a mixture of ``normal'' dust (illuminated by a plausible range of
radiation intensities) plus emission from
a population of small ($a\ltsim0.01\micron$)
magnetic nanoparticles.
We consider three magnetic materials: metallic Fe, magnetite Fe$_3$O$_4$,
and maghemite $\gamma$-Fe$_2$O$_3$. 
It appears that any of these 3 materials, or a combination of them,
can provide enough emission at $\lambda > 1\mm$ so that a combination
of ``normal dust'', spinning dust, and magnetic dust can acount for
the observed SED of the SMC.

\item If conditions in the SMC are conducive to a large fraction of
the interstellar Fe being in magnetic nanoparticles, other low metallicity
galaxies may also have mm-wave emission dominated by magnetic dipole
emission.

\item While it seems natural for the magnetic nanoparticles to
be inclusions in larger grains, the observed extinction does not
rule out the possibility that the magnetic nanoparticles might be
independent free-fliers.

\item If the magnetic nanoparticles are present as randomly-oriented
inclusions in larger silicate grains, the polarization is expected
to fall as the frequency decreases below $\sim200\GHz$.
It may be possible to test this prediction with 
measurements by {\it Planck} of the polarized emission
from the SMC.

\item Our models are based on high-frequency magnetic properties
as estimated by \citet{Draine+Hensley_2012a} using the Gilbert equation.
Laboratory studies of the high-frequency ($\nu\gtsim100\GHz$)
magnetic properties of metallic Fe and Fe oxides are needed to improve
our understanding of magnetism at high frequencies.
\end{enumerate}

\acknowledgements

We thank 
J.-P.\ Bernard,
K.\ Gordon,
S.\ Stanimirovic,
and
G.W.\ Wilson
for helpful discussions regarding the SMC.
This research made use of NASA's Astrophysics Data System Service,
and was supported in part
by NSF grant AST 1008570.
BH acknowledges support from a NSF Graduate Research Fellowship under
Grant No. DGE-0646086.

\bibliography{btdrefs}

\end{document}